# Superconductivity in nanocrystalline tungsten thin films growth by sputtering in a nitrogen-argon mixture

J. A. Hofer,[1*] N. Haberkorn[1,2]

[1] Comisión Nacional de Energía Atómica and Consejo Nacional de Investigaciones Científicas y Técnicas, Centro Atómico Bariloche, Av. Bustillo 9500, 8400 San Carlos de Bariloche, Argentina.

[2] Instituto Balseiro, Universidad Nacional de Cuyo and Comisión Nacional de Energía Atómica, Av. Bustillo 9500, 8400 San Carlos de Bariloche, Argentina.

We report on the structural and superconducting properties of nanocrystalline tungsten thin films growth by sputtering at room temperature with an $N_2$:Ar mixture ($N_2$ from 3% to 50%). The crystalline phases were identified by comparing as-grown and thermal annealed thin films. For $N_2/(Ar+N_2)$ mixtures between 3 and 10 %, the films display nanocrystalline β-W phase. Coexistence of β-W and $W_2N$ phases are observed for gas mixtures with $N_2$ between 20 % and 40%. A detailed study of the superconducting properties as function of the thickness was performed for W films growth with 8 % $N_2$ mixtures. For this concentration, the nitrogen atoms increase the disorder at the nanoscale reducing the grain size and avoiding the crystallization of α-W. The superconducting critical temperature ($T_c$ = 4.7 K) is thickness independent for films thicker than ~ 17 nm. Below this thickness, the $T_c$ value decreases systematically being 3.1 K for 4 nm thick films. Our study provides a simple method for the fabrication of nanocrystalline β-W thin films with potential applications in superconducting devices.

## 1. Introduction

Tungsten is a metal with several applications in micro-electromechanical devices [1,2]. W films exist in two allotropic forms: alpha (α)-W (body-centered cubic structure) and beta (β)-W (A15 structure). The electrical resistivity of α-W is always lower than the electrical resistivity of β-W. A distinctive feature of W is that displays superconductivity [3,4]. The superconducting critical temperature $T_c$ can be enhanced from 12 mK for α-W to up 5.2 K for the coexistence of nanocrystalline β-W and amorphous structures [5,6]. A simple method to study the influence of the microstructure on the electronic properties of nanocrystalline W is growing thin films. Among the different techniques for the development of thin films, sputtering allows the growth of high-quality samples even at a low substrate temperature. The structural, mechanical and electronic properties usually depend on the deposition parameters [7,8,9]. Extremely disordered structures can be achieved by both reducing the deposition temperature and adding impurities. For example, nanocrystalline W thin films with coexistence of β– and α- phases can be obtained by sputtering at room temperature [8]. In addition, nanocrystalline β–W has been stabilized in thin films growth at low temperatures by reactive sputtering with Ar/$O_2$ [10] and Ar/$N_2$ mixtures [9,11]. The high stability of metallic W growth in reactive atmospheres can be related to low heat of formation [12]. While it is clear that the interstitial impurities and the structural disorder stabilize the β-phase, only a few works have discussed their influence on the thermal stability and the superconducting properties [8,11,13].

In this work, we analyze the presence of superconductivity in nanometric grain size W thin films growth at room temperature in an $N_2$/Ar mixture ($N_2$ from 3% to 50%). The results show that the microstructure, crystalline structure and the presence of superconductivity depend on the $N_2$ concentration in the gas mixture. The comparison between as-grown and thermal annealed films shows that the superconductivity can be related to nanocrystalline β-W phase and amorphous phases. We study in detail the superconducting properties for films growth with an 8% $N_2$ mixture. For this concentration, no features related with $W_2N$ were observed for X-ray diffraction in as-grown samples (nitrogen is mainly an interstitial impurity). The study of the superconducting properties was performed for films with thickness between 3 nm and 160 nm. The superconducting critical temperature ($T_c$ = 4.7 K) is thickness independent for films thicker than ~ 17 nm. Below this thickness, the $T_c$ value decreases systematically being 3.1 K for 4 nm thick films. The analysis of the upper critical fields $H_{c2}$ indicates that the extrapolated value at zero increases as the films thickness decreases, which can be related to changes in the microstructure and reduction in the grain size average.

## 2. Materials and methods

W films were deposited on Si (100) wafers by sputtering in an Ar / $N_2$ mixture. The substrates ($\approx$ 1x$10^{-4}$ $m^2$) were etched with HF 6%. The films were grown using a power of 50 watts with a 99.99% purity W target (diameter: 0.038 m). During the deposition, the substrate is positioned directly over the targets at ~0.055 m. The substrate temperature was modified from room temperature (RT) to 673 K. The base pressure was ≈ $6.7\times10^{-5}$ Pa. The total pressure during the deposition of the films was kept constant in 0.67 Pa. The $N_2$ partial pressure was modified from 3% to 50%. Ultra-high purity Ar (99.999%) and $N_2$ (99.999%) were used as gas sources. The deposition rate is ≈ 23 nm/min for $N_2$ concentrations between 3 and 10 % and decreases systematically to ≈ 20 nm/min for gas mixtures with 50 % of $N_2$. Thermal annealing was performed in a vacuum atmosphere (≈$1.33\times10^{-4}$ Pa) at 973 K. Wherever used, the notation [*d*WX-*T*], indicates a tungsten film with a thickness of *d* [nm], and X nitrogen [%] mixture and T the growth temperature [K], respectively.

X-ray diffraction (XRD) was used to characterize the crystal structure of the films (Panalytical Empyrean equipment). The structural analysis was performed based on Θ-2Θ scans under Cu Kα radiation operated at 40 kV and 30 mA using an angular resolution of 0.02°. The film thickness was calibrated by low-angle X-ray reflectivity (XRR). Profile fitting was done using the Parratt32 code [14]. The electrical transport measurements were performed on 1 mm (length) x 0.02 mm (width) bridges using the standard four-terminal transport technique (see sketch in Fig. 1).

## 3. Results and discussion

*3.1. Influence of the $N_2$/Ar mixture on the crystalline structure and electrical transport of W thin films*

Figure 2*a* shows the XRD pattern for W films growth using different $N_2$:Ar mixtures. The results show that for $N_2$ mixtures up 10 %, the XRD patterns display the (200) and (210) reflections of the β phase. For $N_2$ mixtures between 20% and 40%, the (200) reflection is not observed and the (210) is at 2θ ~ 38.5°. The shift in the angle may be related to a long lattice parameter due to higher disorder produced by a large amount of interstitial nitrogen. The XRD pattern for [138W50-RT] displays a peak at 2θ ~ 36.45°, which could be related to the (111) reflection of nanocrystalline $W_2N$ [9]. The grain size average (*D*), extracted from the analysis of the peak width using the Scherrer formula, decreases as the $N_2$ in the mixture (the exception is [138W50-RT]). The estimated *D* values are 9 nm for [160W8-RT], 6.7 nm for [160W10-RT], 2.1 nm for [155W20-RT], 1.8 nm for [145W30-RT] and 7.9 nm for [138W50-RT]. It is important to note that sputtering usually produces columnar structures, which implies that the grains could be elongated at the growth direction. To analyze in detail the origin of the peaks in the XRD patterns, the films were annealed at 973 K for 1 hour in vacuum. Figure 2*b* shows the results for [160W8-RT], [155W20-RT], [145W30-RT] and [138W50-RT]. Thin films growth with $N_2$ concentration up 10% display phase coexistence between $\alpha$-W and $W_2N$ (minority phase), which suggests that the nitrogen in the films growth at RT is an interstitial impurity. The *D* value for $\alpha$-W is $\approx$ 17 nm. Thin films with mixtures between 20 % and 30 % display coexistence of $W_2N$ (*D* $\approx$ 40 nm and 70 nm, respectively) and metallic W (minority phase). Finally, the XRD pattern for [138W50-RT] displays the (111) and (200) reflections of the $W_2N$ phase [9]. The shift to lower angles of both reflections respect to tabulated values can be related to interstitial nitrogen [15]. The *D* value is $\approx$ 12 nm.

Figure 3*a* shows a summary of the electrical transport for thin films growth using different $N_2$ mixtures. The criteria used to determine the $T_c$ is indicated in [138W50-RT]. Most of the films display superconductivity with $T_c$ between 3 K and 4.9 K (see Fig. 3a). The exception is [155W20-RT]. The $T_c$ values are in the range of amorphous W [5,6]. Moreover, for $N_2$ mixtures lower than 10%, the fluctuations in $T_c$ are similar to that reported in ref. [8]. We attribute the presence of superconductivity for $N_2$ mixtures richer than 20% to the coexistence of amorphous W and nitrides. The disorder at the nanoscale for all the films is evident from the low value of the residual resistivity ratio ($RRR = R^{300K}/R^{10K}$) $\approx$ 0.9-0.97 (not shown). Following the analysis of the structural and superconducting properties will be focused on thin films growth using an 8 % $N_2$ mixture, which according to XRD data produces mainly nanocrystalline $\beta$-W.

*3.2. Influence of the deposition temperature on structural and superconducting properties of nanocrystalline β-W thin films*

To analyze the influence of deposition temperature on the microstructure and the electrical transport, 160 nm thick W thin films were grown at different temperatures using an 8% $N_2$ mixture. Figure 4*a* shows the XRD for [160W8-RT], [160W8-473], [160W8-573] and [160W8-673]. All the samples display a peak that can be attributed to the $(210)_\beta$ reflection. The peak is systematically narrow and shifted to higher angles when the deposition temperature is increased. The reduction in the peak width can be related with an increment in the grain size average. The shift in the peak position and the asymmetry to higher angles can be related with crystallization $\alpha$-W and overlapping of the $(210)_\beta$ and $(200)_\alpha$ reflections (see for example [160W8-673]). It is important to note that the asymmetry corresponding to the $(200)_\alpha$ reflection is lesser evidenced at 673 K than 473 K. The differences may be related to an increment in the grain size and a better overlapping due to a reduction of the peak width. Figure 4*b* shows the temperature dependence of the normalized resistance ($R/R^{10K}$) for [160W8-RT], [160W8-473] and [160W8-573]. The results show that $T_c$ decreases systematically as the deposition temperature increases. The drop in $T_c$ could be related to the coexistence of β-W and α-W phases (evidenced by XRD). As we will discuss in the section 3.3, the reduction in the dimension of the superconducting paths decreases the $T_c$ value.

To understand the influence of the thermal annealing on the superconducting properties, we measure the upper critical fields ($H_{c2}$) in [160W8-RT] and [160W8-473]. For polycrystalline films, the anisotropy $\gamma = H_{c2}^{\parallel}/H_{c2}^{\perp}$ is related to the geometry [16]. Figure 4*c* shows a summary of the results. First, we will analyze $H_{c2}^{\perp}(T)$, and second, the differences between $H_{c2}^{\perp}(T)$ and $H_{c2}^{\parallel}(T)$. The temperature dependences of $H_{c2}^{\perp}$ for dirty superconductors can be analyzed by the Werthamer-Helfand-Hohenberg (WHH) formula

$$ln\frac{1}{t} = \sum_{v=-\infty}^{\infty}\left(\frac{1}{|2v+1|} - \left[|2v+1| + \frac{\hbar}{t} + \frac{(\alpha\hbar/t)^2}{|2v+1|+(\hbar+\lambda_{so})/t}\right]^{-1}\right) \text{ [eq. 1]}$$

where $t = T / T_c$, $\hbar = (4/\pi^2)(H_{c2}(T)/|dH_{c2}/dT|_{T_c})$, α is the Maki parameter, and $\lambda_{so}$ is the spin-orbit scattering constant. When $\lambda_{so}$ = 0, $H_{c2}(0)$ obtained from the WHH formula satisfies the relation $H_{c2}(0) = \frac{H_{c2}^{orb}(0)}{\sqrt{1+\alpha^2}}$ [17]. The data for both films is well described by the model using α = 0 and $\lambda_{so}$ = 0 (see dashed lines in Fig. 4*c*). The coherence length ξ (0) values can be estimated using $\xi(0)=\sqrt{\Phi_0/(2\pi H_{c2}^{\parallel}(0))}$ (with $\Phi_0$ = 2.07 x 10$^{-15}$ Wb is the flux quantum). The obtained values using $H_{c2}^{\perp}$ = 4.6 T for [160W8-473] and 6.6 T for [160W8-RT] are 8.5 nm and 7.2 nm, respectively. We now turn to the comparison between $H_{c2}^{\perp}$ and $H_{c2}^{\parallel}$. It is important to mention that both films are in the 3D limit (ξ << 160 nm). The $\gamma = H_{c2}^{\parallel}/H_{c2}^{\perp}$ is ≈ 1.7 for [160W8-RT] and ≈ 2.6 for [160W8-473]. The surface superconductivity produces a field enhancement of $H_{c3}$ = 1.69 $H_{c2}$, which is in agreement with for [160W8-RT] [16]. On the other hand, γ = 2.6 is in agreement with the 2D limit where $H_{c2}^{\parallel}(T) \propto (1-t)^{1/2}$ [16]. The so-called vortex-free state is observed in films thinner than ≈ 4.4ξ (the film is too thin to nucleate a vortex core) [18], which for [160W8-473] implies the presence of superconducting paths thinner than the nominal thickness (reduction of the superconductor volume due to crystallization of α-W). We discuss the features of the 2D behavior later when we analyze the $H_{c2}(T)$ dependences for films with different thickness.

*3.3. Thickness dependence of the structural and superconducting properties of nanocrystalline β-W thin films*

To understand the influence of the dimension on the superconducting properties of nanocrystalline β-W thin films, we growth samples at room temperature using an 8% $N_2$ mixture with a thickness between 3 and 160 nm. The thicknesses were determined from the fits. Figure 5a shows the data for 11.8 nm and 22.8 nm thick W films. It is important to note that the best fits were obtained adding an interfacial layer of approximately 1 nm. This layer was not considered in the nominal thickness value. Figure 5b shows a summary of the thickness obtained from fits versus deposition time. The resulting growth rate average is of 21.0 (0.5) nm/min.

Figure 6 shows the XRD pattern for [11W8-RT], [23W8-RT], [70W8-RT] and [160W8-RT]. The results show reflections of the β phase. The peaks are wider as the thickness reduces, which can be related to size and microstructural effects. The thicknesses of the layers affect the reflections in the XRD (the peak width increases as the thickness reduces). Moreover, higher disorder for the thinner films may be expected due to lower growth temperature average due to self-heating during the sputtering process (shorter growth times).

Figure 7a shows the normalized resistance of W films with different thickness. The results show that films thicker than 17 nm show a $T_c$ = 4.7 K. Below this thickness, the $T_c$ value decreases systematically, being 3.1 K for a 3.8 nm thick W film. The superconducting transition width is $\Delta T_c = T_c - R^{zero} \approx 0.1$ K in most films. A slight broadening of the superconducting transition is observed in the thinner films, which can be attributed to enhanced thermal fluctuations when the thickness falls below the coherence length ξ [16]. The films display residual resistivity ratio ($RRR = R^{300K}/R^{10K}$) ≈ 0.93 – 1.04, which is indicative of high disorder with very short mean free path (not shown). The resistivity of the films is independent of the thickness ρ (300 K) ≈ 1.5 μΩ.m.

Figures 8ab show $H_{c2}(T)$ for [8W8-RT] and [17W8-RT] with the magnetic field **H** perpendicular (⊥) and parallel (∥) to the surface. Inset Fig. 8a shows typical curves of normalized resistance ($R/R^{10K}$) versus temperature for [8W8-RT] with **H** ∥ surface. The graph includes the corresponding fits using the WHH and 2D models, respectively. For the $H_{c2}^{\perp}$ (T) dependences, the $H_{c2}$ (0) values for [8W8-RT] and [17W8-RT] are ≈ 9 T (ξ (0) = 6.0 nm) and 8 T (ξ (0) = 6.4 nm), respectively. The comparison with [160W8-RT] (see Fig. 4c) indicates that the $H_{c2}$ (0) increases as the thickness decreases. This change could be attributed to the reduction in the grain size with thickness discussed above. The increment in the $H_{c2}$ (0) values in nanocrystalline systems reducing the grain size is also evidenced in other systems such as $Mo_2N$ thin films [19,20].

The temperature dependence of $H_{c2}^{\parallel}$ in the thin films with thickness lower than approximately 4.4ξ can be analyzed in the 2D limits:

$$H_{c2}^{\parallel}(T) = \frac{\sqrt{3}\Phi_0}{\pi d \left[0.855(\xi_0 l)^{\frac{1}{2}}\right]} (1 - T/T_c)^{1/2} \quad [\text{eq. 2}]$$

with ξ (0) = (ξ₀l)^½ (l is the mean free path and ξ₀ the coherence length in the clean limit). The equation consider that in the dirty limit ξ(T) = 0.855ξ(0) (1-(T/T_c))^{-1/2} [16]. Although the predicted temperature dependences of superconducting parameters are for T→ T_c, they usually are valid over a much wider temperature range. The fits using eq. 2 for $H_{c2}^{\parallel}$ [8W8-RT] with ξ (0) = 6.0 nm and [17W8-RT] with ξ (0) = 6.4 nm are in good agreement with the experimental data. It is important to note that for [8W8-RT], $H_{c2}^{\parallel}$(2.8K) = 16 T and the extrapolation to zero is above 26 T. These values are above to the Pauli paramagnetic limit $H_p$ ≈ 1.84 $T_c$ (in Tesla for $T_c$ in Kelvin) for isotropic BCS superconductors [21]. Above $H_p$, it is expected that the Zeeman splitting energy matches the superconducting energy gap or binding energy of a Copper pair. The presence of very huge $H_{c2}$ values in the vortex-free configuration suggests that the same tolerance to the magnetic field should be observed in nanocrystalline nanowires.

## 4. Summary

The superconducting and microstructural properties of W thin films fabricated by sputtering in $N_2$:Ar mixtures were studied. The microstructure and phases are strongly affected by the gas mixture. Nanocrystalline β-W phase is obtained for $N_2$ concentration between 3% and 10 %. Thin films growth at RT using $N_2$ concentration between 20 % and 40 % display coexistence of β-W and $W_2N$ phases. It is important to note that for $N_2$ concentration up 10 %, the nitrogen stabilizes the β-W phase. This behavior is different to other metals such as tantalum [11] and molybdenum [19] where reactive mixtures of Ar/$N_2$ produce nitrides.

The study of the superconducting properties is focused in nanocrystalline β-W thin films growth at RT using 8% $N_2$ mixtures. Coexistence of β-W and α-W phases can be induced by both thermal annealing and increment in the deposition temperature. The phase coexistence reduces the superconductor volume. The superconducting critical temperature ($T_c$ = 4.7 K) is thickness independent for films thicker than ~ 17 nm. The $H_{c2}^{\perp}$ (T) dependences are well described by the WHH model for single band superconductors in the dirty limit. The extrapolated $H_{c2}$ (0) values increase as the thickness reduces, which can be related to changes in the microstructure (small grain sizes for thinner films). On the other hand, the $H_{c2}^{\parallel}(T)$ dependences, depending on thickness, are well described considering surface superconductivity and the 2D model. Our study provides a simple method for the fabrication of nanocrystalline β-W with potential applications in superconducting devices.

**Acknowledgements**

This work has been partially supported by Agencia Nacional de Promoción Científica y Tecnológica PICT 2015–2171 and CONICET PIP 2015-0100575CO. JAH and NH are members of the Instituto de Nanociencia y Nanotecnología CNEA-CONICET (Argentina).

Figure 1. Schematic of 4-point probe configuration. V and I represent voltage and current contacts, respectively.

Figure 2. XRD patterns for: *a)* As-grown [160W8-RT], [160W10-RT], [155W20-RT], [145W30-RT] and [138W50-RT]. *b)* Same samples annealed at 973 K. The peak deconvolution considering $Cu_{K\alpha 1}$ and $Cu_{K\alpha 1}$ for [138W50RT] is included.

Figure 3. *a)* Temperature dependence of the normalized resistance ($R/R^{10K}$) for W thin films growth using different $N_2$/Ar gas mixtures. The criteria used to determine the $T_c$ is indicated in [138W50-RT]. *b)* Summary of $T_c$ (right) and residual resistivity ratio ($RRR$ =$R^{300K}/ R^{10K}$) (left).

Figure 4. *a)* XRD patterns for W thin films growth in 8% N2 concentration at different temperatures. *b)* Temperature dependence of the normalized resistance ($R/R^{10K}$). *c)* Temperature dependence of the upper critical field ($H_{c2}$) with **H** parallel and perpendicular to the surface for [160W8-RT] and [160W8-473].

Figure 5. *a)* XRR patterns for W-thin films growth during 0.5 and 1 minute. *b)* Summary of the film thickness resulting from fits as function of the deposition time.

Figure 6. XRD patterns for [11W8-RT], [23W8-RT], [70W8-RT] and [160W8-RT].

Figure 7. *a)* Normalized resistivity as a function of temperature for W thin films with different thickness. *b)* Summary of $T_c$ versus thickness. Inset shows a zoom for thickness lower than 20 nm. The data for 2.8 nm is extrapolated from the temperature dependence of $R/R^{10K}$.

Figure 8. *a-b*) Temperature dependence of the upper critical field ($H_{c2}$) with **H** parallel (∥) and perpendicular (⊥) to the surface for [8W8-RT] and [17W8-17], respectively. Inset *a)* shows typical curves of normalized resistance ($R/R^{10K}$) versus temperature for [8W8-RT] with **H** ∥ surface (from right to left $\mu_0 H$ = 0, 8 T, 10 T, 12 T and 16 T).

Figure 1.

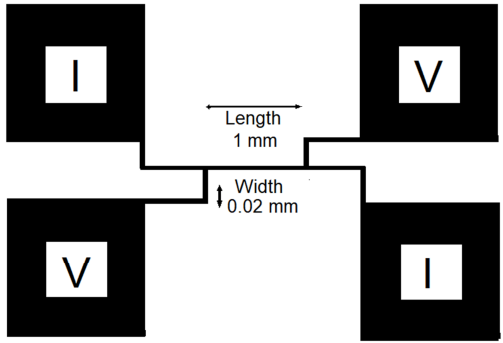

Figure 2.

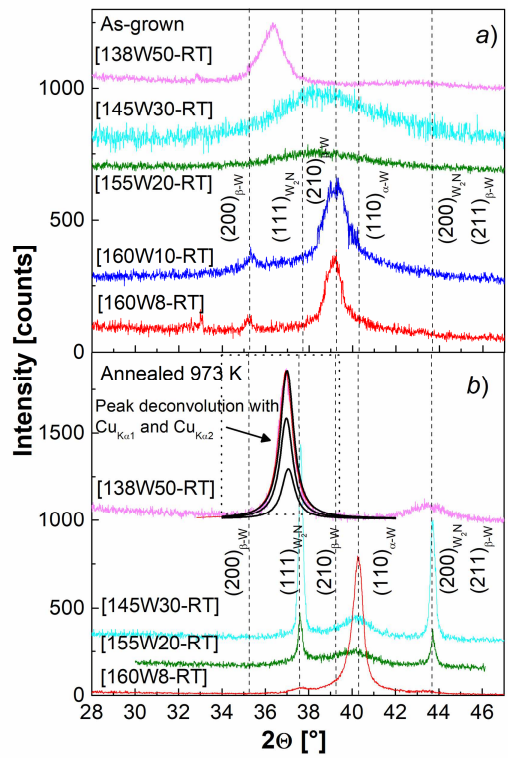

Figure 3.

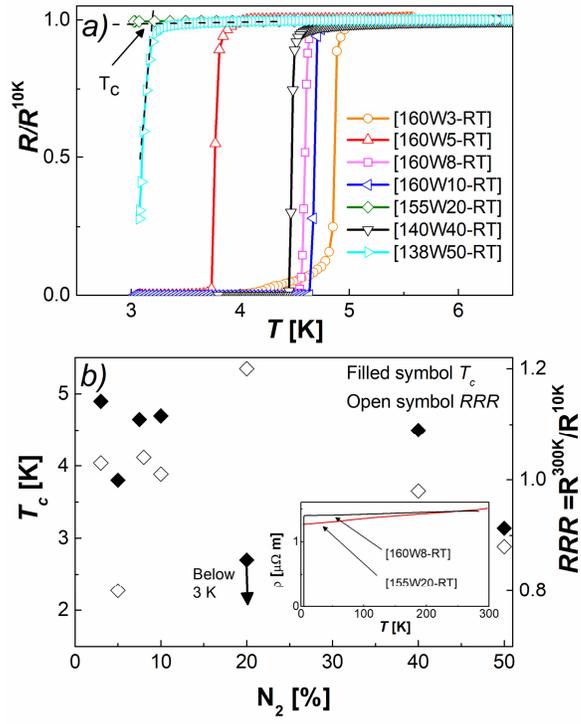

Figure 4.

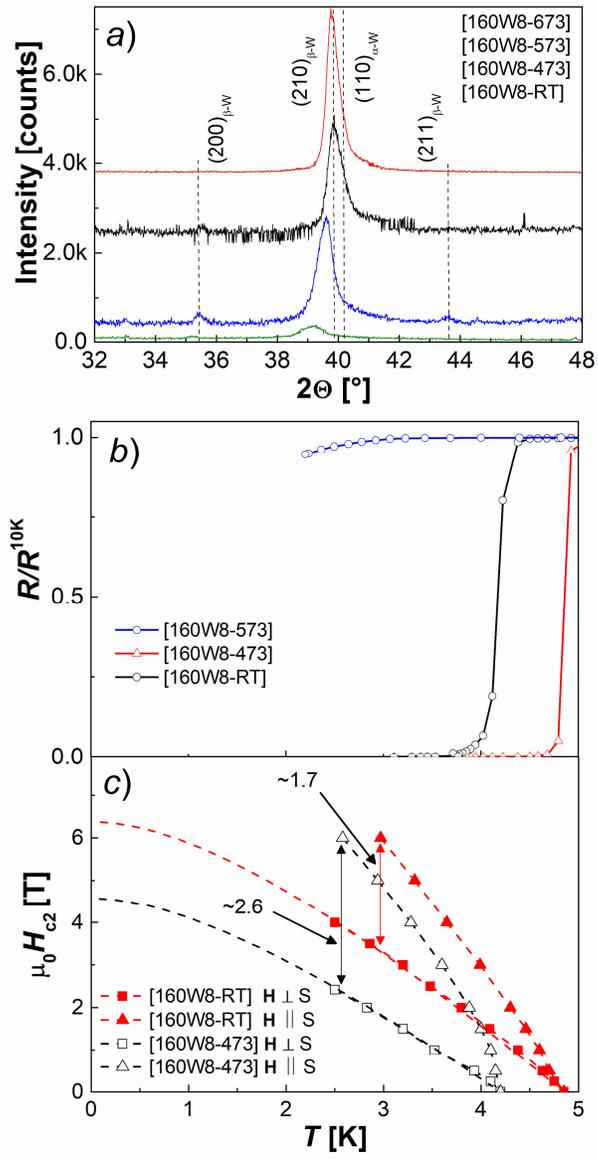

Figure 5.

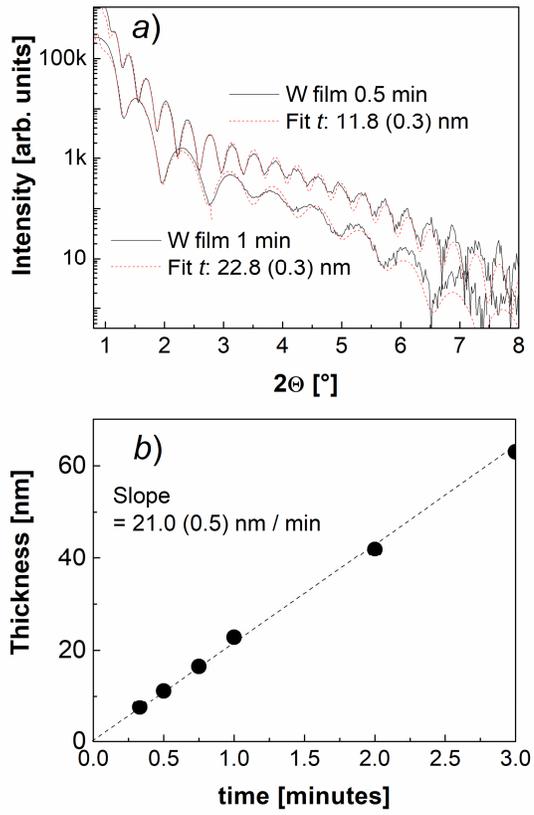

Figure 6.

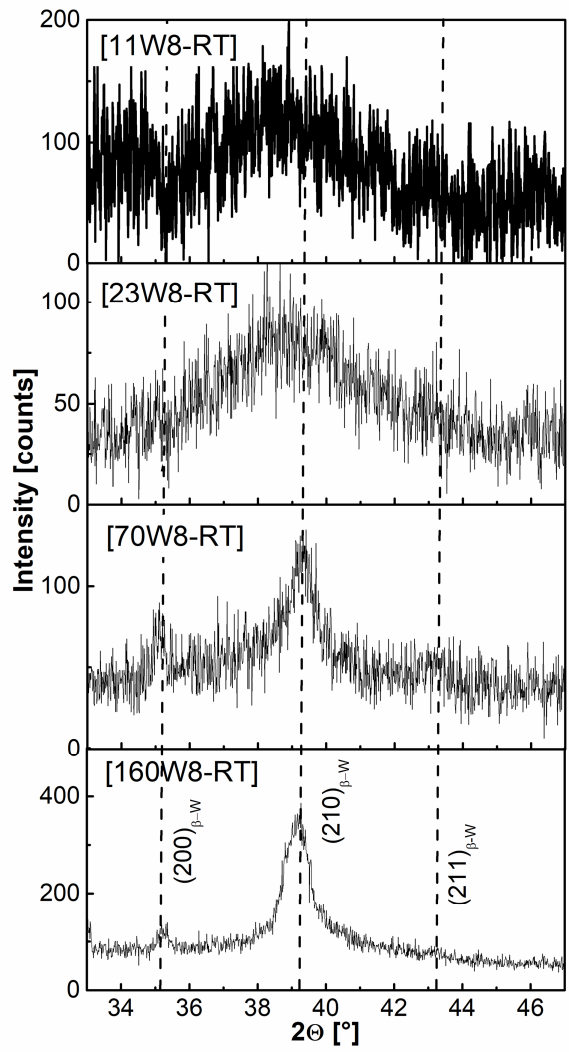

Figure 7.

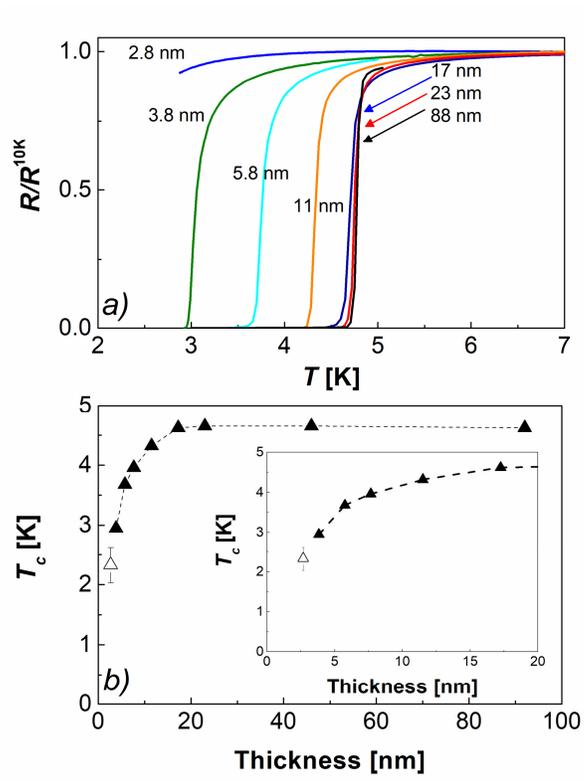

Figure 8.

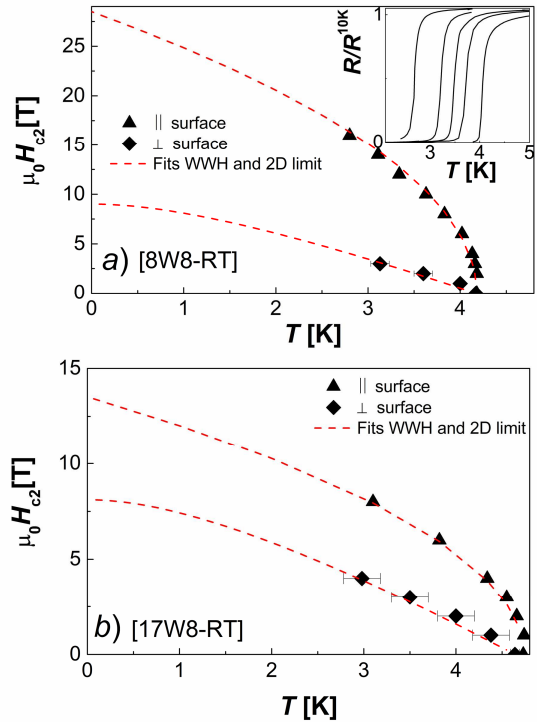